\documentclass[times, 10pt,twocolumn]{article} 
\usepackage{latex8}
\usepackage{times}
\usepackage{url}
\usepackage{graphicx}
\usepackage{amssymb}

\usepackage{color}
\newcommand{\YY}[1]{{\color{black}#1}}

\begin{document}

\title{ Learning to Predict on Octree for Scalable Point Cloud Geometry Coding \vspace{-12pt}}
\author{
Yixiang Mao, Yueyu Hu, Yao Wang\\
Department of Electrical and Computer Engineering \\ 
New York University, Brooklyn, NY 11201, USA\\ 
\{yixiang.mao, yyhu, yw523\}@nyu.edu\\
\vspace{-16pt}}

\maketitle
\thispagestyle{empty}

\begin{abstract}
Octree-based point cloud representation and compression have been adopted by the MPEG G-PCC standard. However, it only uses handcrafted methods to predict the probability that a leaf node is non-empty, which is then used for entropy coding. We propose a novel approach for  predicting such probabilities for geometry coding, which applies a denoising neural network to a ``noisy'' context cube that includes both neighboring decoded voxels as well as uncoded  voxels. We further propose a convolution-based model to upsample the decoded point cloud at a coarse resolution on the decoder side. Integration of the two approaches significantly improves the rate-distortion performance for geometry coding  compared to the original G-PCC standard and other baseline methods for dense point clouds. The proposed octree-based entropy coding approach is naturally scalable, which is desirable for dynamic rate adaptation in  point cloud streaming systems.
\end{abstract}

\vspace{-12pt}
\Section{Introduction}
\vspace{-6pt}

Efficient coding of point cloud data is critical for the continued development of lifelike virtual reality (VR) experiences.
For geometry-based point cloud compression, in order to overcome the memory inefficiency from using uniform voxel grids, the 3D space is typically recursively subdivided into smaller cubes using octree where only non-empty nodes are further subdivided \cite{meagher1982geometric}.
The octree coding mode in in the MPEG G-PCC standard uses a handcrafted context table for context-based entropy coding \cite{TMC13_doc}.

We propose a novel entropy coding method for the octree-based geometry coding.
For each non-empty parent node, to predict the probability that each of its 8 children nodes is occupied , we  form an initial ``noise'' context cube. For example, if we use a context that includes $5\times 5 \times 5$ parent nodes ($k=5$), the context cube will include $10\times 10 \times 10$ children nodes.   The  nodes that have been coded in the context cube will have context values of either 1 (occupied) or 0 (empty),   nodes that have not been  coded will have context values of 0 if they correspond to empty parent nodes, and finally the uncoded nodes that correspond to occupied parent nodes will be assigned a value of 0.5. We then apply a 3D convolution-based neural network to denoise the values in the cube and use the output values of the denoised cube at the center $2\times 2 \times 2$ voxels as the predicted probability. As with G-PCC, we code the non-empty nodes from the top level of the octree to the next level, sequentially, naturally yielding a scalable bit-stream.

On the decoder side, if the received bit-stream only includes a partial representation of the full octree, corresponding to  a coarse (lossy) representation of the original point cloud,  we further propose a 3D convolution-based neural network to upsample the reconstructed point cloud. We have found that such upsampling at the decoder side can significantly improve the quality of the reconstructed point cloud.

In addition to fully convolutional models, we also develop  alternative probability estimation and upsampling models with significantly less complexity while maintaining comparable coding efficiency. 
We train our proposed and baseline models on a subset of ShapeNet \cite{chang2015shapenet}, and evaluate the performance on the 8iVSLF dataset recommended by MPEG \cite{krivokuca20188i}.
Compared with G-PCC \cite{TMC13_doc} (using a handcrafted entropy model) and VoxContext-Net \cite{que2021voxelcontext} (using a machine learning method), our method saves around 80\% bits for achieving the same reconstruction quality during lossy compression (by stopping before the final octree level), or saves around 30\% bits when the point cloud is losslessly coded.

A key advantage of our work from the prior related studies is that our octree-based entropy coding model is naturally scalable. The bitstream can be organized into segments, where each segment corresponds to an octree level. Hence, decoding an octree level only requires the information from the earlier part of the bitstream. This scalable coding setting benefits the future design of streaming systems, enabling the streaming system to dynamically change the delivery rate based on the channel conditions. It would also enable the streaming systems to perform intelligent prefetching and correction. For example, the system may  prefetch future video segments at a lower rate to prevent freeze, and  fetch additional bits to enhance the quality of the prefetched version at a later time when more bandwidth is available.

\vspace{-4pt}
\Section{Related works}
\vspace{-6pt}

The point cloud compression (PCC) methods in the literature can be categorized into two classes: video-based (V-PCC) and geometry-based (G-PCC) \cite{graziosi2020overview}.
Video-based methods usually first generate 3D surface segments by dividing the point cloud into some connected regions, called 3D patches. Then, each 3D patch is projected independently into a 2D plane, and those patches on the 2D plane are organized and coded by traditional video encoders. Video-based methods are heavily investigated and the performance benefits from the well-developed 2D video encoders. MPEG-PCC already released video-based PCC standard (ISO/IEC 23090-5) in 2021 \cite{2021vpcc}.
Meanwhile, geometry-based methods encode the coordinates and colors of the points directly in 3D space.
Geometry-based methods are developing fast in recent years and experts are exploring both traditional and deep-learning methods.

For the handcrafted geometry-based methods,  tree structures are usually used (e.g., octree \cite{meagher1982geometric} or KD-tree \cite{devillers2000geometric}) to recursively divide the 3D space.
Since the first work that uses the octree to present the 3D geometry  \cite{meagher1982geometric}, more traditional methods \cite{peng2003octree,garcia2018intra,huang2008generic,kammerl2012real,schnabel2006octree} emerged using octree variants or considering temporal information to further improve the coding efficiency.
The MPEG group has been  developing a geometry-based PCC standard (G-PCC) \cite{2021gpcc}  using a handcrafted entropy model,  and its corresponding test model (TMC13) \cite{TMC13,TMC13_doc} has been made available and supports the octree coding mode. Another widely-used open-source G-PCC software, Google’s Draco, uses a KD-tree compression method \cite{Google_Draco}.

Several prior studies also exploited geometry-based point cloud compression using deep learning models. 
Some works construct the uniform voxel grids of the entire 3D space to represent the point cloud \cite{maturana2015voxnet,huang20193d,yang2018pixor,quach2019learning,zhou2018voxelnet}. Those works perform the 3D convolution on the voxel grids and consume an extensive amount of memory space to save the voxel grids, which leads to the inefficiency of processing large or sparse point cloud data. A recent work \cite{wang2021multiscale} performs sparse convolution on the voxel grids to reduce the time and space complexity.
Another work \cite{guarda2020adaptive} improves the performance of G-PCC by both spatially and temporally adapting to the optimal deep learning entropy model based on the characteristics of the point cloud. However, switching between entropy models makes this method not scalable. Previous study \cite{quach2020improved} introduces a set of improvements to the entropy model and training strategy to achieve a lower rate but the codec is also not scalable.

While the MPEG group is working on standardizing the G-PCC, using the octree is recognized as a good approach  with several benefits (e.g., saving memory consumption and being scalable). 
Recently, several studies apply the deep learning model on the octree nodes \cite{huang2020octsqueeze,que2021voxelcontext} that shows strong potential for performing 3D convolution while still using octree structure.
To predict the  probability for the leaf nodes, the earlier work \cite{huang2020octsqueeze} forms the node's context only from the node's ancestors (parent nodes); the later work, VoxContext-Net (VCN) \cite{que2021voxelcontext} uses the context from the spatial neighbors in the previous coded level (one level above).
However, none of them use the strong context information from the currently coded octree level. Additionally, the coordinate refine model proposed in VCN \cite{que2021voxelcontext} only adjusts the node location, which does not realize the full potential of post-processing at the decoder side.

In this work, we propose to use the context information from the currently coded octree level for the entropy coding model; we further propose to use the upsampling as the post-processing step after lossy decoding. Both approaches significantly improve the coding performance.


\vspace{-4pt}
\Section{Proposed Methods}
\vspace{-6pt}

\SubSection{Context-Based Entropy Coding for Octree Geometry: Notation and Basic Ideas}
\vspace{-6pt}

\begin{figure}
  \includegraphics[width=1\linewidth]{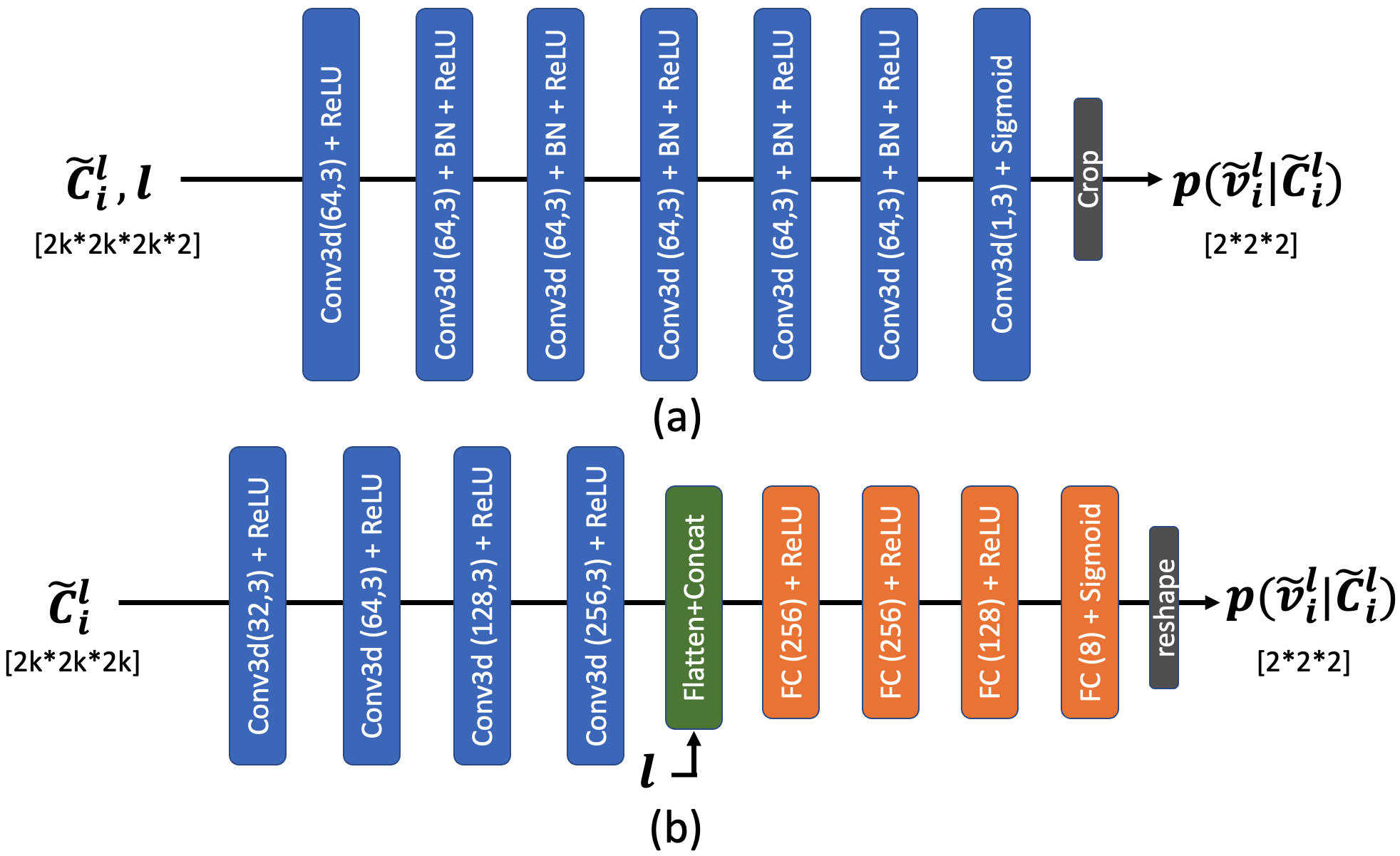}
  \vspace{-20pt}
  \caption{Network architectures for entropy coding: (a) Model A: the proposed convolution only architecture, (b) Model B: the convolution + fully connected architecture. 
  ``Conv(n,l)" means a 3D convolutional layer outputting $n$ feature channels  with a 3D filter of kernel size of $l\times l \times l$.}
  \vspace{-22pt}
  \label{fig:entropy_network}
\end{figure}


\YY{

To achieve scalability, we adopt the octree representation. A point cloud is  represented in a sequence of $L$ occupancy  levels: $\mathbf{X}_1, \mathbf{X}_2, \cdots, \mathbf{X}_L$. The octree is coded from the first level occupancy $\mathbf{X}_1$ to the last level $\mathbf{X}_L$, sequentially and losslessly.  The point cloud can be reconstructed to level $l$ if the coded bit-streams for $\mathbf{X}_1, \mathbf{X}_2, \cdots, \mathbf{X}_l$ are given.  When the coded bit-streams for all occupancy levels are given, we can losslessly reconstruct the original point cloud.

Each non-empty node $x_{l-1, i}$ at level $l-1$, location $i$ has 8 children at level $l$, represented in occupancy $\tilde{v}_i^l = \{v_j \in \{0,1\}: j=1,2,\cdots,8\}$. 
With context-based entropy coding, we code  $\tilde{v}_i^l$ into a bit-stream  based on the conditional probability mass function $p(\tilde{v}_i^l|{\tilde C}_i^l)$, where  ${\tilde C}_i^l$ represents the context. In general ${\tilde C}_i^l$ may include nodes in  $\mathbf{X}_1, \mathbf{X}_2, \cdots, \mathbf{X}_{l-1}$, as well as the adjacent nodes in $\mathbf{X}_l$ that have been coded. With the same context available at the decoder, the decoding process is to map the bit-stream back to $\tilde{v}_i^l$. 

}
\vspace{-6pt}
\SubSection{\YY{Conditional Probability Estimation through Denoising  a ``Noisy'' Context Cube}}
\vspace{-6pt}

The main challenge in  context-based entropy coding is how to form the context ${\tilde C}_i^l)$ and how to estimate the probability distribution $p(\tilde{v}_i^l|{\tilde C}_i^l)$.
To accurately estimate the probability distribution, it is important to design a context that fully utilizes all the  information from previously decoded nodes. When coding an octree node at the current level $\mathbf{X}_l$, the context should take into account of both the information at the upper level $\mathbf{X}_{l-1}$ and the current level. However, since not all the information at the current level is known when the current node is being coded, we need to treat the known and the unknown part separately to ensure that the context for probability estimation used during the encoding is available during the decoding process.

In the proposed probability estimation method, we code the nodes $ \tilde{v}_i^l$ following a fixed spatial order. We ensure that when a node at the spatial position $(x_i,y_i,z_i)$ is being coded, nodes at the same level with coordinates $c \in \{(x,y,z):z<z_i\} \cup \{(x,y,z):y<y_i, z=z_i\} \cup \{(x,y,z):x<x_i,y=y_i,z=z_i\}$ have already been coded. When coding $\tilde{v}_i^l$, we form a context cube ${\tilde C}_i^l$ of size $2k\times 2k \times 2k$ centered at $ \tilde{v}_i^l$, corresponding to a cube of size $k\times k \times k$ at the level $l-1$. The known half of the context cube is filled with the true occupancy, \textit{i.e.} 0 for empty voxels and 1 for occupied ones. For the other half of ${\tilde C}_i^l$ that has not been coded, we fill them with 0 if their parent nodes  indicate that these children nodes are empty, and 0.5 if their parent nodes indicate that the unknown voxel can be potentially non-empty. Since this context involves uncertainty in signal values in the uncoded voxels, we call it ``noisy'' context. We  propose to use a convolution neural network to ``denoise'' this context, so that the output of the network represents $p(\tilde{v}_i^l|{\tilde C}_i^l)$, the predicted probabilities that the center $2\times 2\times 2$ voxels are non-empty. These predicted probabilities will then be used by an entropy coder.

To reduce the complexity,  we train one neural network  for probability estimation at all octree {levels}. To make use of the level information, along with the original noisy context we add another 3D input channel with the same size of $2k\times 2k\times 2k$, and all elements in this channel are set to the level index $l$. The resulting  3D tensor provides the information about the neighboring voxels' occupancy (which is noisy for the unknown half)  and the level in the octree.  
%
%
Inspired by the architecture of DnCNN for image denoising \cite{zhang2017beyond}, we design the network architecture shown in Fig.~\ref{fig:entropy_network}(a). The network takes the context cube  and the octree level channel as input. \YY{A sigmoid activation is used at the final layer to generate the probability distribution $\mathbf{p} \in \mathbb{R}^8$, with each element $p_j \in [0,1], j=1,2,\cdots,8$, where $p_j := Pr\{v_j = 1\}$.}

Compared to the network used in VoxelContext-Net~\cite{que2021voxelcontext}, this network does not have fully connected layers at the end in order to preserve more spatial information. 
We also develop another probability estimation network with fully connected layers, similar to the one used in the VoxelContext-Net~\cite{que2021voxelcontext}, shown in Fig.~\ref{fig:entropy_network}(b). However, the context  in \cite{que2021voxelcontext} only uses the occupancy information in the parent level. We will provide  performance comparison of these two network architectures in Sec.~\ref{sec:ab_reslt}.

\vspace{-6pt}
\SubSection{\YY{Resolution Enhancement of Decoded Point Clouds}}
\vspace{-6pt}
\begin{figure}
  \includegraphics[width=1\linewidth]{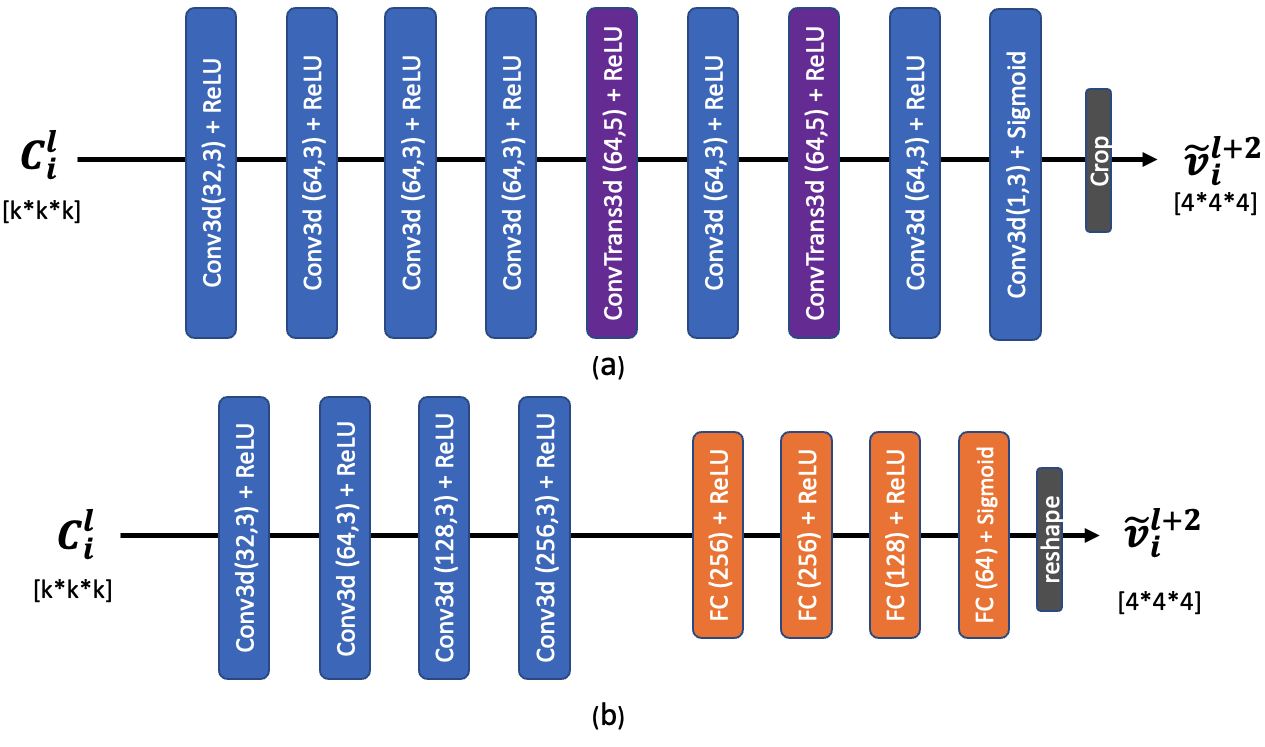}
  \vspace{-20pt}
  \caption{Network architectures for upsampling: (a) the proposed convolution-only architecture, (b) the convolution + fully connected architecture. 
  }
  \vspace{-22pt}
  \label{fig:post_network}
\end{figure}

\YY{When the bit-rate is restricted due to the network throughput constraint, the point cloud cannot be transmitted in full precision. With the proposed scalable coding method, the bitstream may only contain information up to level $l$ of the octree. To further enhance the quality of the reconstructed point cloud, we propose to estimate a finer resolution representation of the point cloud at the decoder side.} After losslessly decoding the bitstream to the $l$-th level of the octree, we upsample each octree node at the $l$-th level, to a $4\times4\times4$ \YY{voxel grid}, to generate a lossy reconstruction of the original octree up to level $l+2$ without using additional bits. 

\YY{We upsample a node based on its neighboring context.} For a node at location $i$ from the decoded $l$-th level point cloud, a local voxel context $C_i^l$ centered at this node is formed.
\YY{The network takes $C_i^l$ as the input and maps it to $\tilde{v}_i^{l+2}$ that is  4 times larger along each dimension than the input.} The center $4\times4\times4$ voxels of the $C_i^{l+2}$ are then binarized  and used to form the upsampled point cloud at level $l+2$. This model is repeatedly applied to each node at the $l$-th level without conditional dependency. Thus, the upsampling of all nodes can run simultaneously on a multi-threaded processing unit (e.g., a GPU) to vastly reduce the processing time.

\YY{The network architecture is shown in Fig.~\ref{fig:post_network}~(a). The network takes an input tensor with size of $k \times k \times k$, and maps it to a tensor with size  $4k \times 4k \times 4k$. From the output tensor, the center $4\times 4\times 4$ cube is cropped, and binarized with a threshold $t$ to the final predicted occupancy voxel grid.}

\YY{Since the density and pattern of points at different levels on the octree are diverse,}
we train the upsampling network separately for every depth level.
Note that the model upsamples the octree by two levels only when the decoded point cloud level $l \leq L-2$, where $L$ is the maximal level of the original point cloud. When $l=L-1$, a similar model with only one upsampling layer is used to estimate the decoded point cloud from the $l$-th level to the full levels. When $l=L$ (lossless coding), no post-processing model is applied.

\YY{For comparison,} we also develop an alternative architecture with fully connected layers at the end to predict the upsampled points, similar to the one used in the VoxelContext-Net~\cite{que2021voxelcontext}, as shown in Fig.~\ref{fig:post_network}~(b). The performance comparison is provided in Sec.~\ref{sec:ab_reslt}.

\vspace{-6pt}
\SubSection{Loss Function}
\vspace{-6pt}

\YY{We train the probability estimation network to directly minimize the expected bits needed to code the occupancy.
This is equal to the binary cross entropy~(BCE) loss function over the ground truth occupancy and the predicted probability. The loss for each training sample (corresponding to one non-empty parent node) is
\begin{equation}
    \mathcal{L}_e(\mathbf{x}, \mathbf{q})  = - \sum_{j=1}^{8} x_j \log q_j + (1-x_j) \log (1 - q_j),
\end{equation}
where $x_j\in \{0,1\}$ denotes the occupancy ground truth and $q_j$ is the estimated probability, $q_j = Pr\{v_j = 1\}$.

We adopt the same loss function for the training of the upsampling network, which upsamples the decoded point cloud from level $l$ to level $l+2$ (when $l<L-1$). The loss function is calculated between the ground truth and the predicted probabilities both at the target upsampled level.}



\vspace{-4pt}
\Section{Experiments}
\vspace{-6pt}

\SubSection{Experimental Setup}
\vspace{-8pt}

\subsubsection{Datasets}
\vspace{-4pt}
\YY{We train the network based on point clouds sampled from the ShapeNetCore~\cite{chang2015shapenet} dataset.
The dataset consists of a total number of 51,300 3D object models in 55 categories, each with mesh and texture.}
We randomly choose 1024 objects across all categories from ShapeNetCore, \YY{and densely sample point clouds on the mesh given by the models. The coordinates of the sampled points are quantized to a bit-depth of 10, with duplicate points removed. We build voxel grids and octrees of depth 10 on the point clouds. The voxel grids and octrees are used to train the proposed networks.}

To ensure that our trained networks  generalizes to other dense point clouds besides the simple objects in the training set, we evaluate the proposed method on the 8i Voxelized Surface Light Field (8iVSLF) dataset~\cite{krivokuca20188i}, which is recommended by MPEG for dense point cloud coding experiments.
We compare with two baseline methods, G-PCC and VCN,  on the four 10-bit point cloud frames: \textit{longdress\_vox10\_1300}, \textit{loot\_vox10\_1200}, \textit{redandblack\_vox10\_1550}, and \textit{soldier\_vox10\_0690}.

\begin{table}[]
\begin{center}
\caption{Bits per point (bpp) used for losslessly coding 8i dense point clouds, by MPEG G-PCC,  VoxContextNet(VCN), and our Model A and our Model B. (VCN,  Model A and  Model B all use $k=5$)}
\label{tab:rate}
\scalebox{0.9}{
\begin{tabular}{|l|c|c|c|c|}
\hline
Models         & G-PCC & VCN & Model A & Model B \\ \hline
Longdress & 1.02  &   1.25   & 0.67   &    0.72   \\ 
Loot & 0.95  &     1.23    & 0.65     &    0.69 \\ 
Redandblack & 1.08  &  1.31    & 0.77   &   0.82    \\ 
Soldier & 1.01  &   1.28   & 0.69   &   0.74    \\ \hline
\textbf{Average (bpp)} & 1.02  & 1.26  & 0.69    &   0.74   \\ \hline
\textbf{Rate reduction}     & 0 &    +23.5\%   & -32.1\%   &   -27.3\%   \\ \hline
\end{tabular}}
\end{center}
\vspace{-16pt}
\end{table}


\vspace{-8pt}
\subsubsection{Evaluation Metric}
\vspace{-4pt}

We measure the quality of the reconstructed point cloud with the point-to-point (D1) PSNR~\cite{tian2017geometric, mekuria2017performance}, calculated using the MPEG PCC DMetrics software~\cite{DMetrics}.  The bit-rate is given in bit-per-point (bpp), calculated by dividing the number of bits over the total point number in the original point cloud.
\vspace{-8pt}
\subsubsection{Baseline Methods}
\vspace{-4pt}
The first baseline method is MPEG standard Geometry Point Cloud Compression (G-PCC) \cite{TMC13_doc}. We follow the common test conditions (CTC) \cite{graziosi2020overview} to generate the baseline G-PCC Rate-distortion (R-D) curve. 
We use the TMC13 \cite{TMC13} and enable the G-PCC octree codec to code the dense point cloud. 
In order to generate the R-D points for variable bit rates, we set the G-PCC to code the octree level by level and truncate at different levels.

The second baseline method is VoxelContext-Net \cite{que2021voxelcontext}. In this work, the author did both training and testing on the same ScanNet dataset, which may lead to model overfitting to the specific dataset and its performance on the MPEG G-PCC standard dataset is unclear.
To fairly compare the performance, we train this baseline method on the same subset of the ShapeNetCore dataset and test the it on 8iVSLF dataset.
Since there is no publicly available source code of VoxelContext-Net, the training code is reproduced by ourselves.

\vspace{-6pt}
\SubSection{Experiment Results}
\vspace{-6pt}

\label{sec:exp_reslt}

\begin{figure}
  \includegraphics[width=1.05\linewidth]{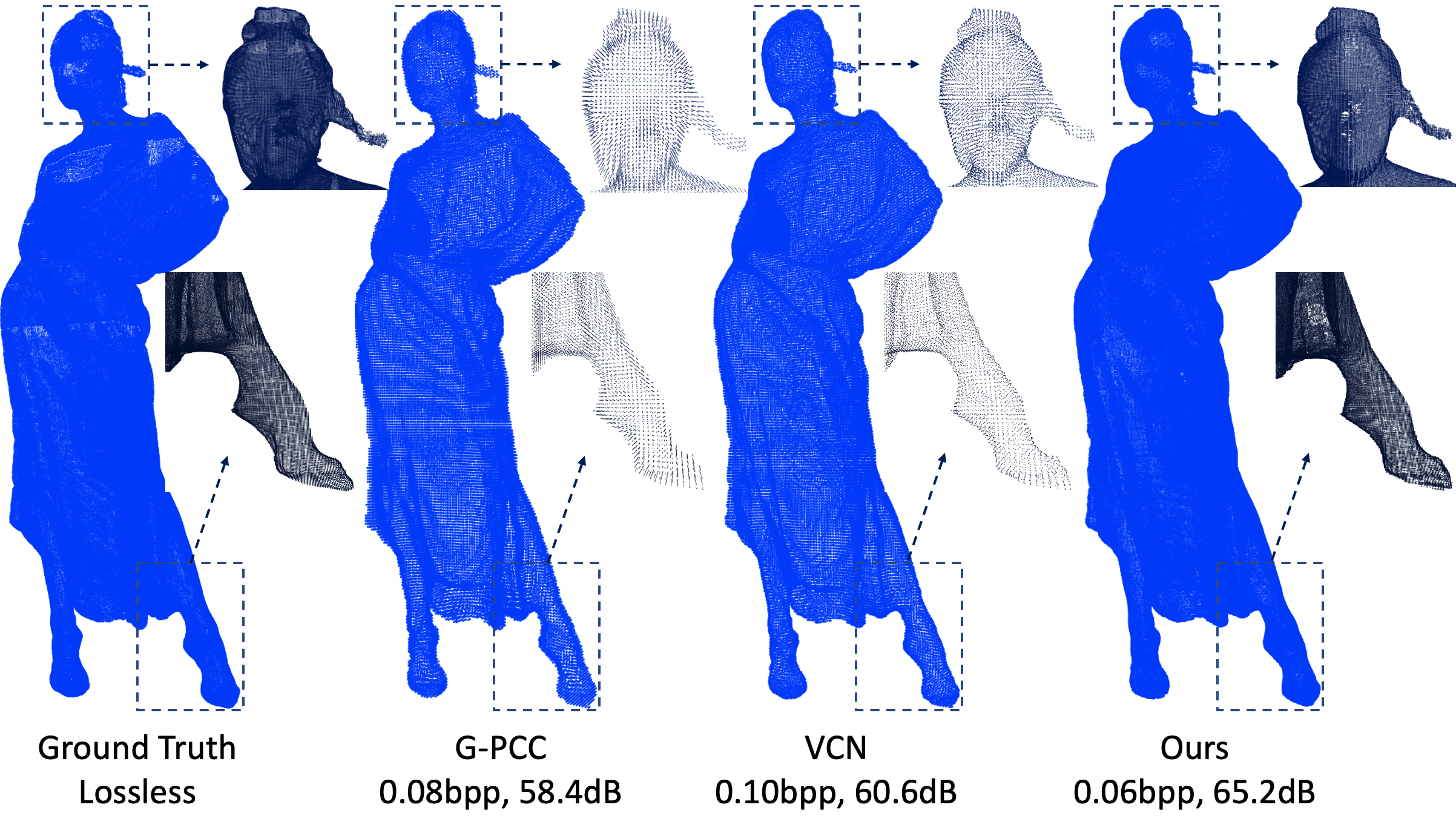}
  \vspace{-16pt}
  \caption{Visualization of the ground truth, G-PCC, VCN, and ours. 
\vspace{-20pt}
  }
  \label{fig:visual}
\end{figure}


For lossless compression, we calculate the bpp for G-PCC, VoxContext-net and the proposed method, on the test point clouds, the results are shown in Table~\ref{tab:rate}. For the proposed method, we show the results using both Model A,  shown in Fig.~\ref{fig:entropy_network}(a), and  Model B,  shown in Fig.~\ref{fig:entropy_network}(b). The context size is $k=5$.  Note that the upsampling model is not used for the lossless compression. As shown, Model A and Model B both outperform the baseline methods.
Compared to G-PCC, Model A and Model B reduce the number of bits by 32.1\% and 27.0\%, respectively,  on average. For the VoxContext-net model, we observe that it requires more bits than G-PCC to compress the dense 8iVSLF point cloud. This differs from the performance gain over G-PCC reported in \cite{que2021voxelcontext}. This could be caused by several reasons: Firstly, we train the model on a subset of the ShapeNet dataset and test on the 8iVSLF dataset for a fair comparison, whereas the original paper \cite{que2021voxelcontext} reports the testing results  on the ShapeNet dataset; Secondly, we used only a subset of the ShapeNet dataset for training whereas the work in \cite{que2021voxelcontext} used the entire training set of the ShapeNet; Finally, the handcrafted G-PCC explores the previously coded neighbors in the current coding level while VoxContext-net does not, and such information could be especially useful for the dense point cloud data. 

The rate-distortion (RD) curves of lossy compression by these methods are shown in Fig.~\ref{fig:modelvs}. 
Compared with G-PCC and VoxContextNet, both our Model A and Model B save around 80\% of bit rate. Model A uses the fully convolutional network in Fig.~\ref{fig:entropy_network}(a) for probability prediction, and the fully convolutional network in Fig.~\ref{fig:post_network}(a) for upsampling. Model B uses the architectures in Fig.~\ref{fig:entropy_network}(b) and  Fig.~\ref{fig:post_network}(b), for probability prediction and upsampling, respectively.   
Visualization results of \textit{longdress\_vox10\_1300} for our proposed method and baseline methods are shown in Fig.~\ref{fig:visual}.

\vspace{-6pt}
\subsection{Ablation Study}
\vspace{-6pt}
\label{sec:ab_reslt}

\subsubsection{\YY{Network Architecture}}
\vspace{-4pt}
\begin{figure}
  \begin{center}
  \includegraphics[width=1\linewidth]{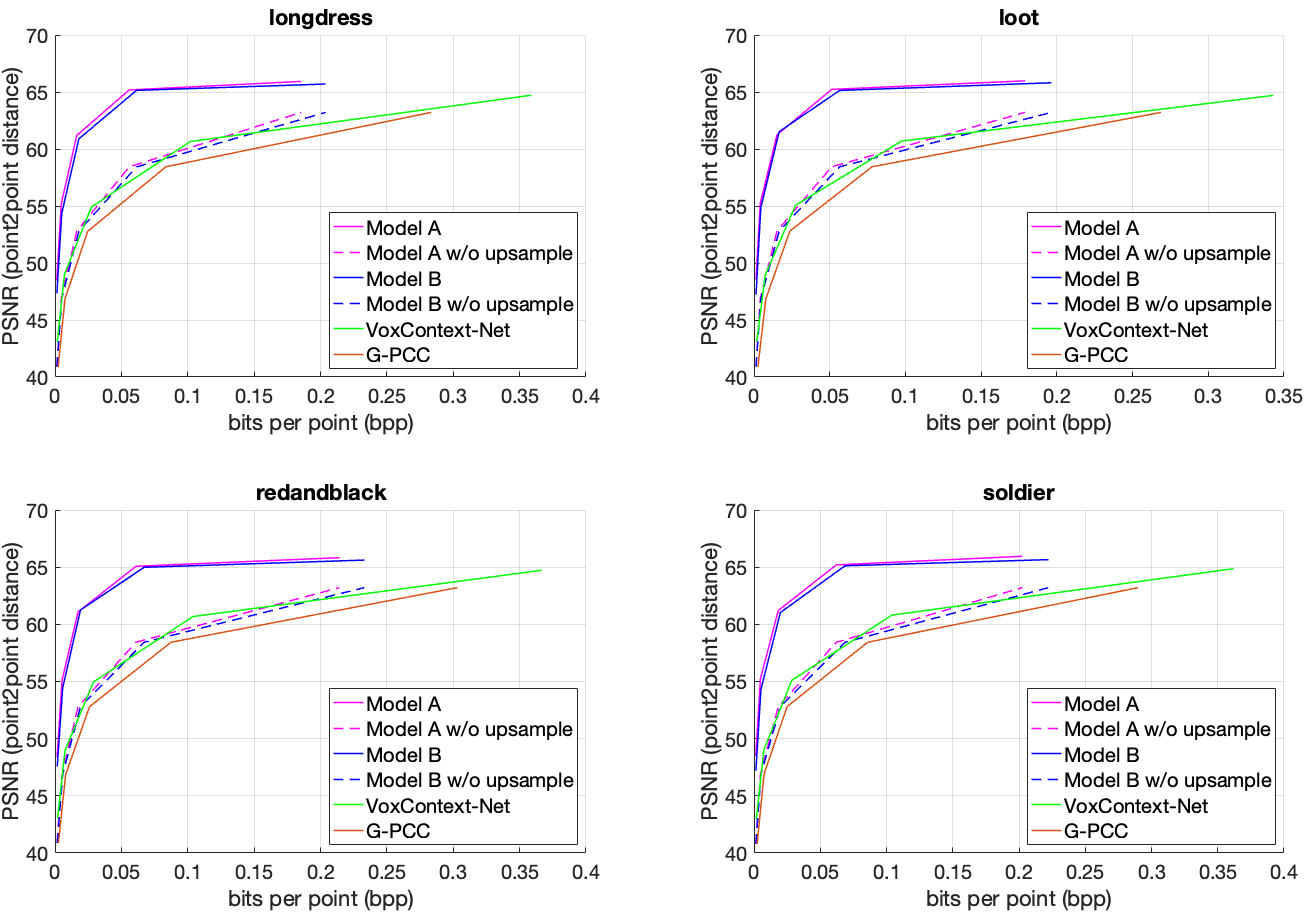}
  \end{center}
  \vspace{-16pt}
  \caption{Model A (fully convolutional network) vs. Model B (convolutional+fully connected network).}
  \vspace{-22pt}
  \label{fig:modelvs}
\end{figure}

\begin{table}[]
\centering
\caption{Number  of floating point operations (KFLOPS per node)
for the probability estimation (E) and the post-processing (P) procedures, respectively, with different methods and at different  voxel context  sizes for the Longdress sequence. B(3) means the proposed Model B using context size with $k=3$.}
\begin{tabular}{|c|c|c|c|c|c|}
\hline
&  VCN(5) &   A(5)   &   B(3)  &   B(5)  &   B(9)  \\ \hline
E &  93.2  &  559.7  &  30.1  &  30.4  & 206.9  \\ \hline
P &  149.4  &  404.9  &  33.2  &  60.4  &  73.0  \\ \hline
\end{tabular}
\label{tab:flops}
\end{table}



For both the probability estimation and  point cloud upsampling tasks, the architecture with fully-connected layers (Model B) is slightly less efficient in terms of rate-distortion trade-off  than the fully convolutional one (Model A), as shown in Fig.~\ref{fig:modelvs}.
However, Model B significantly reduces the computational complexity. Table~\ref{tab:flops} compares the number of floating point operations (FLOPS) of these two models. As shown,  inference using Model B takes only about 15\% of the FLOPS compared to Model A.
Given that the coding efficiency of using Model B does not drop significantly compared to Model A, Model B may be preferred for practical applications.

\vspace{-10pt}
\subsubsection{Upsampling Strategies}
\vspace{-4pt}
\begin{figure}
  \begin{center}
  \includegraphics[width=0.75\linewidth]{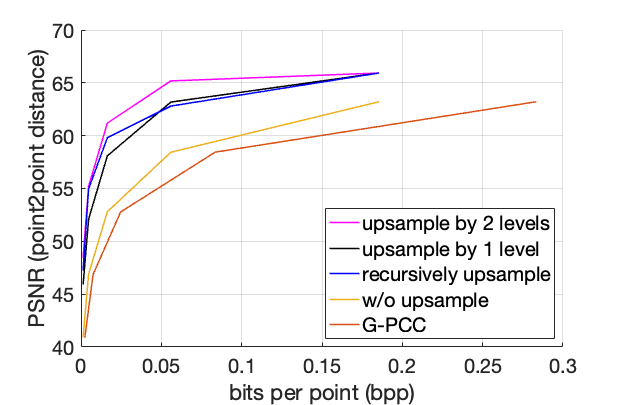}
  \end{center}
  \vspace{-16pt}
  \caption{Comparison between different postprocessing (upsampling) strategies. For example, for the  upsampling by 2 levels curve,  the highest rate point is obtained by coding the octree to level 8 and upsample to level 10; while  the next highest rate point is obtained by coding the octree to level 7 and upsample to level 9.}
  \vspace{-22pt}
  \label{fig:upsamplevs}
\end{figure}

\YY{In this ablation study, we compare the proposed scheme with two alternative upsampling strategies: 1) upsample by one level only; 2) recursively upsample by one level until the final level (different models are used at different levels). Fig.~\ref{fig:upsamplevs} compares their R-D performances . The experiments are conducted on \textit{longdress\_vox10\_1300}}.
Compared with upsampling by one level, the model that directly upsamples by two levels has consistent and significant gain in reconstruction quality. However, recursively upsampling by one level sometimes leads to worser quality. This is because the upsampling error at an earlier level propagates, and negatively affects the upsampling for the following levels. Therefore, we adopt the two-level upsampling strategy.

\vspace{-10pt}
\subsubsection{Context Cube Size}
\vspace{-4pt}
\YY{The dimensionality of the context cube affects the probability estimation accuracy and  upsampling accuracy. To determine the best size of the context, }
we train Model B using different context sizes, and compare their R-D performances on the \textit{longdress\_vox10\_1300} point cloud in Fig.~\ref{fig:upsamplevs}. 
The performance gain of using a larger context cube diminishes when the context size $k$ exceeds 5 (corresponding to a context cube of   $10\times10\times10$ for probability estimation, and $5\times 5 \times 5$ for upsampling).  The FLOPS of the  models with different context sizes are given in Table~\ref{tab:flops}.  To balance the performance and the complexity, $k=5$ is the preferred choice and is used in the results shown in Fig.~\ref{fig:modelvs}, Fig.~\ref{fig:upsamplevs} and Table~\ref{tab:flops} .
For  complexity-sensitive applications, a context  size of $k=3$ could also be used, which only suffer from slight degradation in rate-distortion performance, as shown in Fig.~\ref{fig:upsamplevs}.

\begin{figure}
  \begin{center}
  \includegraphics[width=0.75\linewidth]{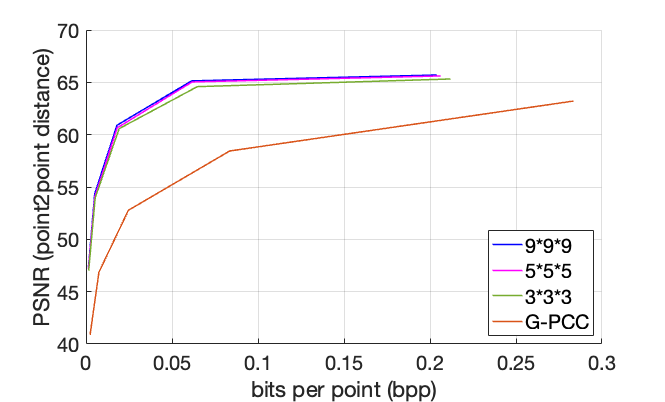}
  \end{center}
  \vspace{-16pt}
  \caption{R-D performances corresponding to different context sizes. The experiments are conducted with Model B.}
  \vspace{-22pt}
  \label{fig:upsamplevs}
\end{figure}

\vspace{-4pt}
\Section{Conclusions}
\vspace{-6pt}
In this paper, we propose an octree-based point cloud geometry compression method using machine learning models. Our first contribution considers context-based entropy coding of octree nodes. We  form a ``noisy'' context using the occupancy information at the currently coded octree level, and use 3D convolution-based ``denoising" networks to predict the probability that an octree node is non-empty. The second contribution considers decoder post-processing and proposes  convolutional networks to upsample a low-resolution point cloud (corresponding to a low bit rate resulting from coding to a low level of the octree) to a higher resolution (corresponding to a higher octree level).   The combination of the probability estimation and the upsampling approaches significantly improves the rate-distortion performance of octree-based geometry coding over the current MPEG standard G-PCC as well as several prior works leveraging machine learning. We further  compare different network  structures for both the probability estimation task and the upsampling task in terms of both the rate-distortion performance and computational complexity.   Being an octree-based geometry coding solution, our method naturally leads to a scalable bit stream and has strong potential to be adopted in future point cloud streaming platforms.

\bibliographystyle{latex8}
\bibliography{latex8}

\begin{thebibliography}{10}\setlength{\itemsep}{-1ex}\small

\bibitem{Google_Draco}
{Google Draco}.
\newblock \url{https://github.com/google/draco}.

\bibitem{TMC13}
{MPEG G-PCC TMC13}.
\newblock \url{https://github.com/MPEGGroup/mpeg-pcc-tmc13}.

\bibitem{DMetrics}
{MPEG PCC DMetrics}.
\newblock
  \url{http://mpegx.int-evry.fr/software/MPEG/PCC/mpeg-pcc-dmetric.git}.

\bibitem{TMC13_doc}
{MPEG Point Cloud Compression}.
\newblock
  \url{https://mpeg-pcc.org/index.php/public-contributions/g-pcc-codec-description}.

\bibitem{2021vpcc}
Information technology--coded representation of immersive media -- part 5:
  Visual volumetric video-based coding (v3c) and video-based point cloud
  compression (v-pcc).
\newblock {\em ISO/IEC}, pages 23090--5, 2021.

\bibitem{2021gpcc}
Information technology — coded representation of immersive media — part 9:
  Geometry-based point cloud compression.
\newblock {\em ISO/IEC}, pages 23090--9, Under development.

\bibitem{chang2015shapenet}
A.~X. Chang, T.~Funkhouser, L.~Guibas, P.~Hanrahan, Q.~Huang, Z.~Li,
  S.~Savarese, M.~Savva, S.~Song, H.~Su, et~al.
\newblock Shapenet: An information-rich 3d model repository.
\newblock {\em arXiv preprint arXiv:1512.03012}, 2015.

\bibitem{devillers2000geometric}
O.~Devillers and P.-M. Gandoin.
\newblock Geometric compression for interactive transmission.
\newblock In {\em Proceedings Visualization 2000. VIS 2000 (Cat. No.
  00CH37145)}, pages 319--326. IEEE, 2000.

\bibitem{garcia2018intra}
D.~C. Garcia and R.~L. de~Queiroz.
\newblock Intra-frame context-based octree coding for point-cloud geometry.
\newblock In {\em 2018 25th IEEE International Conference on Image Processing
  (ICIP)}, pages 1807--1811. IEEE, 2018.

\bibitem{graziosi2020overview}
D.~Graziosi, O.~Nakagami, S.~Kuma, A.~Zaghetto, T.~Suzuki, and A.~Tabatabai.
\newblock An overview of ongoing point cloud compression standardization
  activities: Video-based (v-pcc) and geometry-based (g-pcc).
\newblock {\em APSIPA Transactions on Signal and Information Processing}, 9,
  2020.

\bibitem{guarda2020adaptive}
A.~F. Guarda, N.~M. Rodrigues, and F.~Pereira.
\newblock Adaptive deep learning-based point cloud geometry coding.
\newblock {\em IEEE Journal of Selected Topics in Signal Processing},
  15(2):415--430, 2020.

\bibitem{huang2020octsqueeze}
L.~Huang, S.~Wang, K.~Wong, J.~Liu, and R.~Urtasun.
\newblock Octsqueeze: Octree-structured entropy model for lidar compression.
\newblock In {\em Proceedings of the IEEE/CVF conference on computer vision and
  pattern recognition}, pages 1313--1323, 2020.

\bibitem{huang20193d}
T.~Huang and Y.~Liu.
\newblock 3d point cloud geometry compression on deep learning.
\newblock In {\em Proceedings of the 27th ACM international conference on
  multimedia}, pages 890--898, 2019.

\bibitem{huang2008generic}
Y.~Huang, J.~Peng, C.-C.~J. Kuo, and M.~Gopi.
\newblock A generic scheme for progressive point cloud coding.
\newblock {\em IEEE Transactions on Visualization and Computer Graphics},
  14(2):440--453, 2008.

\bibitem{kammerl2012real}
J.~Kammerl, N.~Blodow, R.~B. Rusu, S.~Gedikli, M.~Beetz, and E.~Steinbach.
\newblock Real-time compression of point cloud streams.
\newblock In {\em 2012 IEEE International Conference on Robotics and
  Automation}, pages 778--785. IEEE, 2012.

\bibitem{krivokuca20188i}
M.~Krivokuca, P.~A. Chou, and P.~Savill.
\newblock 8i voxelized surface light field (8ivslf) dataset.
\newblock {\em ISO/IEC JTC1/SC29/WG11 MPEG, input document m42914}, 2018.

\bibitem{maturana2015voxnet}
D.~Maturana and S.~Scherer.
\newblock Voxnet: A 3d convolutional neural network for real-time object
  recognition.
\newblock In {\em 2015 IEEE/RSJ International Conference on Intelligent Robots
  and Systems (IROS)}, pages 922--928. IEEE, 2015.

\bibitem{meagher1982geometric}
D.~Meagher.
\newblock Geometric modeling using octree encoding.
\newblock {\em Computer graphics and image processing}, 19(2):129--147, 1982.

\bibitem{mekuria2017performance}
R.~Mekuria, S.~Laserre, and C.~Tulvan.
\newblock Performance assessment of point cloud compression.
\newblock In {\em 2017 IEEE Visual Communications and Image Processing (VCIP)},
  pages 1--4. IEEE, 2017.

\bibitem{peng2003octree}
J.~Peng and C.~J. Kuo.
\newblock Octree-based progressive geometry encoder.
\newblock In {\em Internet Multimedia Management Systems IV}, volume 5242,
  pages 301--311. SPIE, 2003.

\bibitem{quach2019learning}
M.~Quach, G.~Valenzise, and F.~Dufaux.
\newblock Learning convolutional transforms for lossy point cloud geometry
  compression.
\newblock In {\em 2019 IEEE international conference on image processing
  (ICIP)}, pages 4320--4324. IEEE, 2019.

\bibitem{quach2020improved}
M.~Quach, G.~Valenzise, and F.~Dufaux.
\newblock Improved deep point cloud geometry compression.
\newblock In {\em 2020 IEEE 22nd International Workshop on Multimedia Signal
  Processing (MMSP)}, pages 1--6. IEEE, 2020.

\bibitem{que2021voxelcontext}
Z.~Que, G.~Lu, and D.~Xu.
\newblock Voxelcontext-net: An octree based framework for point cloud
  compression.
\newblock In {\em Proceedings of the IEEE/CVF Conference on Computer Vision and
  Pattern Recognition}, pages 6042--6051, 2021.

\bibitem{schnabel2006octree}
R.~Schnabel and R.~Klein.
\newblock Octree-based point-cloud compression.
\newblock In {\em PBG@ SIGGRAPH}, pages 111--120, 2006.

\bibitem{tian2017geometric}
D.~Tian, H.~Ochimizu, C.~Feng, R.~Cohen, and A.~Vetro.
\newblock Geometric distortion metrics for point cloud compression.
\newblock In {\em 2017 IEEE International Conference on Image Processing
  (ICIP)}, pages 3460--3464. IEEE, 2017.

\bibitem{wang2021multiscale}
J.~Wang, D.~Ding, Z.~Li, and Z.~Ma.
\newblock Multiscale point cloud geometry compression.
\newblock In {\em 2021 Data Compression Conference (DCC)}, pages 73--82. IEEE,
  2021.

\bibitem{yang2018pixor}
B.~Yang, W.~Luo, and R.~Urtasun.
\newblock Pixor: Real-time 3d object detection from point clouds.
\newblock In {\em Proceedings of the IEEE conference on Computer Vision and
  Pattern Recognition}, pages 7652--7660, 2018.

\bibitem{zhang2017beyond}
K.~Zhang, W.~Zuo, Y.~Chen, D.~Meng, and L.~Zhang.
\newblock Beyond a gaussian denoiser: Residual learning of deep cnn for image
  denoising.
\newblock {\em IEEE transactions on image processing}, 26(7):3142--3155, 2017.

\bibitem{zhou2018voxelnet}
Y.~Zhou and O.~Tuzel.
\newblock Voxelnet: End-to-end learning for point cloud based 3d object
  detection.
\newblock In {\em Proceedings of the IEEE conference on computer vision and
  pattern recognition}, pages 4490--4499, 2018.

\end{thebibliography}

\end{document}